\documentclass[amsthm]{elsart}
\usepackage{amsfonts}
\usepackage{amsmath}
\usepackage{graphicx}

\newcommand{\p}{^{\prime}}

\renewcommand{\d}{\,\mathrm{d}}

\newcommand{\ksix}{\xi_{X}}
\newcommand{\ksiy}{\xi_{Y}}
\newcommand{\etax}{\eta_{X}}
\newcommand{\etay}{\eta_{Y}}
\newcommand{\zetap}{\zeta_{+}}
\newcommand{\zetam}{\zeta_{-}}
\newcommand{\zetapm}{\zeta_{\pm}}
\begin{document}
\begin{frontmatter}
\title{Exact solution of the standard transfer problem in a stellar atmosphere}
\runtitle{Exact Solution of the Standard Transfer Problem in a stellar atmosphere}
\date{6 August 2003}
\author{L. Chevallier\thanksref{coresp}}
\and
\author{B. Rutily}
\address{Transfer Group, Centre de Recherche Astronomique de Lyon (UMR 5574 du CNRS),
Observatoire de Lyon, 9 avenue Charles Andr\'{e}, 69561 Saint-Genis-Laval Cedex, France}
\thanks[coresp]{Corresponding author. Tel.:+33-478-868-370; Fax.:+33-478-868-386.\\
\textit{Email address:} \texttt{loic.chevallier@obs.univ-lyon1.fr} (L. Chevallier).}
\runauthor{Chevallier and Rutily}
\begin{abstract}
We come back to the analytical solution of the standard transfer problem in a stellar atmosphere. It consists in solving the radiative transfer equation in a homogeneous and isothermal plane-parallel atmosphere, with light scattering taken as isotropic and monochromatic. The literature on the subject is reviewed and the existing solution in a finite slab is improved thanks to the introduction of non classical auxiliary functions. Eleven-figure tables of the solution are given for typical values of the input parameters currently met in stellar atmospheres.
\end{abstract}
\begin{keyword}
Radiative transfer; Plane-parallel geometry; Isotropic scattering; Stellar atmospheres.
\end{keyword}
\end{frontmatter}

\maketitle

\section{Introduction}\label{sec1}
We consider a static and stationary plane-parallel atmosphere with isotropic and monochromatic scattering in every volume unit.
The source function $S$ of the radiation field is solution to an integral equation of the form (Ivanov~\cite{ivanov1973}, Chap. 3)
\begin{equation}
\label{eq1}
S(\tau) = S^{*} (\tau) + \frac{a(\tau)}{2} \int_{0}^{b} E_{1} (| \tau - \tau\p |) S(\tau\p) \d\tau\p.
\end{equation}
The first term in the right-hand side describes the contribution of the (internal and external) sources, and the second term describes the scattering of photons.
In the latter, $a(\tau) \in [0,1]$ is the volumic albedo for single scattering, and $E_\mathrm{1}$ is the first exponential integral function as defined by
\begin{equation}
\label{eq2}
E_{1} (\tau) = \int_{0}^{1} \exp (-\tau / u) \frac{\d u}{u} \qquad (\tau > 0).
\end{equation}
The optical depth variable $\tau$ covers the range $[0, b]$, where $b>0$ denotes the optical thickness -- possibly infinite --  of the atmosphere at the considered frequency.
Frequency dependence of any quantity is not mentioned.

Suppose that the material is in local thermodynamic equilibrium and that the incoming boundary conditions vanish. Then the source term $S^{*}(\tau)$ is given by Kirchhoff's relation
\begin{equation}
\label{eq3}
S^{*} (\tau) = [1-a(\tau)] B[T(\tau)],
\end{equation}
where $B$ is the Planck function for the temperature $T(\tau)$. \\

The purpose of the present paper is to solve analytically the problem (\ref{eq1})-(\ref{eq3}) for constant $a$ and $T$, that is in a homogeneous and isothermal atmosphere. Then $S^{*} = (1-a) B(T)$ is a constant and the solution reads
\begin{equation}
\label{eq4}
S(\tau) = Q(a,b,\tau) (1-a) B(T) = S(a, b,\tau) B(T),
\end{equation}
where the $Q$- and $S$-functions satisfy the following integral equations: 
\begin{align}
Q(a,b,\tau) &= 1 + \frac{a}{2} \int_{0}^{b} E_{1} (|\tau - \tau\p|) Q(a,b,\tau\p) \d\tau\p, \label{eq5} \\
S(a,b,\tau) &= 1-a + \frac{a}{2} \int_{0}^{b} E_{1} (|\tau - \tau\p|) S(a,b,\tau\p
) \d\tau\p. \label{eq6}
\end{align}
Of course $S=(1-a) Q$ and the above problems are equivalent. In the present paper, we chose to solve Eq. (\ref{eq5}) analytically and to tabulate the function $S$, since the $Q$-function diverges like $1/\sqrt{1-a}$ when $a \to 1$ and $b \to +\infty$. 

We remind the probabilistic meaning of the $Q$-function: $Q(a,b,\tau)$ is the mean number of scatterings experienced by a photon emitted on the $\tau$-layer, which therefore has the probability $P = 1 - (1-a) Q$ to leave the atmosphere since $1-a$ is the photon destruction probability per scattering~\cite{sobolev1967}.
The function $S = (1-a) Q$ is the source function, normalized to the Planck function, of the radiation field propagated in a homogeneous and isothermal atmosphere in local thermodynamic equilibrium.

It is important to solve Eqs. (\ref{eq5}) or (\ref{eq6}) very accurately for different values of $a \in [0,1]$ and $b > 0$, in order to have a benchmark solution available to the validation of codes solving Eq. (\ref{eq1}) numerically.

The literature on the solution of problem (\ref{eq5}) is difficult to follow; it will be seen further that it is incomplete in the finite case ($b < +\infty$).
Several analytical approaches exist, which all rest on a good knowledge of some classical auxiliary functions, firstly the functions $H$, $X$ and $Y$ and their moments.
We remind the definition and useful properties of these functions in Section \ref{sec2}.
Then two classical approaches for solving Eq. (\ref{eq5}) are described. The former is based on the analytical calculation of Sobolev's resolvent function $\phi$ (Section \ref{sec3}), and the latter on the calculation of Ambartsumian's $B$-function (Sec. \ref{sec4}). 
We will see in Section. \ref{sec5} and \ref{sec6} that it is appropriate to express the finite slab solution in terms of non classical auxiliary functions, which we introduce.
Finally, the function $S=(1-a) Q$ is tabulated for some typical values of the parameters $a$ and $b$ in stellar atmospheres (Section \ref{sec7}). 
\section{A reminder on the classical $H$-, $X$- and $Y$-functions}\label{sec2}
The analytical solution to the transfer equation involves a number of auxiliary functions, defined most of time from integral equations modelling the multiple scattering of photons.
The functions $H$, $X$ and $Y$, whose definition is reminded below, are certainly the most famous functions in plane-parallel geometry.
Their detailed study can be found in the classical monographs on radiative transfer theory: Chandrasekhar~\cite{chandrasekhar1950}, Busbridge~\cite{busbridge1960}, $\hdots$, Van de Hulst~\cite{vandehulst1980}, Lenoble~\cite{lenoble1985}.
\subsection{Function $H$}
In a semi-infinite atmosphere ($b = +\infty$), the basic auxiliary function is the $H$-function: it depends on the parameter $a$ and on a angular variable $u$ which we suppose positive.
It can be defined as the unique solution, continuous over $[0, +\infty[$, of either of the following equivalent equations:
\begin{equation}
\label{eq7}
\frac{1}{H(a,u)} = 1 - \frac{a}{2} u \int_{0}^{1} H(a,v) \frac{\d v}{v+u} \, ,
\end{equation}
or
\begin{equation}
\label{eq8}
T(a,u) H(a,u) = 1 + \frac{a}{2} u \int_{0}^{1} H(a,v) \frac{\d v}{v-u}.
\end{equation}

These integral equations on $[0,1]$ define the extension of the $H$-function outside this interval. In Eq. (\ref{eq8}), we have introduced the dispersion function
\begin{equation}
\label{eq9}
T(a,u) = 1 - \frac{a}{2} u \ln \left|\frac{1+u}{1- u} \right| \qquad (u \neq \pm 1),
\end{equation}
and the integral is a Cauchy principal value when $u \in ]0, 1[$.

Equations (\ref{eq7})-(\ref{eq8}) can be solved analytically, a possible expression for their solution being given below [Eq. (\ref{eq61})]. Putting $u \to +\infty$ into (\ref{eq7}) and (\ref{eq8}), and observing that $T(a,+\infty) = 1-a$, one obtains the value at infinity of the $H$-function
\begin{equation}
\label{eq10}
H(a,+\infty) = \frac{1}{\sqrt{1-a}}.
\end{equation}

Busbridge~\cite{busbridge1955} has demonstrated that the $H$-function is solution to the following integral equation:
\begin{equation}
\label{eq11}
H(a, u) = 1 + \frac{R(a)}{H[a, 1/k(a)]} \frac{k(a) u}{1+k(a) u}
 + \frac{a}{2} u \int_{0}^{1} \frac{g(a, v)}{H(a, v)} \frac{\d v}{v + u}.
\end{equation}

For $0<a<1$, the coefficient $k(a)$ is the unique root in $]0,1[$ of the characteristic equation $T \left(a, 1/k(a) \right) = 0$, where $T(a, u)$ is defined by (\ref{eq9}). We thus have by definition
\begin{equation}
\label{eq12}
1-\frac{a}{2}\frac {1}{k(a)}\ln \left[ \frac {1+k(a)}{1-k(a)} \right] = 0 \quad (0<a<1),
\end{equation}
and $k(0)=1$, $k(1)=0$ by continuity.
This coefficient is of great importance when solving the radiative transfer equation in a slab, since $1/k(a)$ is interpreted as the thermalization depth of the atmosphere (Ivanov~\cite{ivanov1973}, Section 3.2).
A complete study of the $k$-function can be found in Case et al.~\cite{case1953}, where $k(a)$ is denoted by $\kappa_\mathrm{0}(c)$.
An analytical expression will be given in Section \ref{sec6}, Eqs. (\ref{eq59})-(\ref{eq60}).

Functions $R(a)$ and $g(a, v)$ appearing in the right-hand side of Eq. (\ref{eq11}) follow immediately from $k(a)$ and $T(a, v)$ since
\begin{equation}
\label{eq13}
R(a) = \frac{1 - k^{2} (a)}{k^{2} (a) + a - 1} \quad(0<a<1)\, ,
\end{equation}
\begin{equation}
\label{eq14}
g(a, v) = \frac{1}{T^{2} (a, v) + [(\pi/2) a v]^{2}} \qquad (0 \leq v < 1).
\end{equation}

Equation (\ref{eq11}) is an integral equation on $[0,1]$, which allows to calculate the $H$-function over $[0, +\infty[$ once it has been solved. Putting $u \to +\infty$ on both sides of Eq. (\ref{eq11}) and taking Eq. (\ref{eq10}) into account yields
\begin{equation}
\label{eq15}
\frac{1}{\sqrt{1-a}} = 1 + \frac{R(a)}{H[a, +1/k(a)]} + \frac{a}{2} \int_{0}^{1} \frac{g(a, v)}{H(a, v)} \d v.
\end{equation}

It will be seen that this equation and its counterpart in a finite slab play a central role in the solution of Eq. (\ref{eq5}).

\subsection{Functions $X$ and $Y$}
In a finite slab ($b < +\infty$), the function $H$ is replaced by two functions, denoted by $X$ and $Y$, which depend on parameters $a$, $b$ and variable $u > 0$. These functions are the unique solution, continuous over $]0, +\infty[$, of either of the following couples of integral equations:
\begin{equation}
\label{eq16}
X(a, b, u) [1 - \ksix (a, b, u)] + Y(a, b, u) \ksiy (a, b, u) = 1,
\end{equation}
\begin{equation}
\label{eq17}
X(a, b, u) \ksiy (a, b, - u) + Y(a, b, u) [1 - \ksix (a, b, - u)] =  \exp (-b / u) \quad (u \neq 1),\end{equation}
or
\begin{equation}
\label{eq18}
T(a, u) X(a, b, u) = 1 - \ksix (a, b, - u) - \ksiy (a, b, u) \exp (-b / u) \quad (u \neq 0, 1),
\end{equation}
\begin{equation}
\label{eq19}
T(a, u) Y(a, b, u) = [1 - \ksix (a, b, u)] \exp (-b / u) - \ksiy (a, b, - u) \quad (u \neq 0, 1).
\end{equation}

We have introduced the functions 
\begin{equation}
\label{eq20}
\ksix (a, b, u) = \frac{a}{2} u \int_{0}^{1} X(a, b, v) \frac{\d v}{v + u} \, ,
\end{equation}
\begin{equation}
\label{eq21}
\ksiy (a, b, u) = \frac{a}{2} u \int_{0}^{1} Y(a, b, v) \frac{\d v}{v + u} \, ,
\end{equation}
defined everywhere except at $u = -1$.
The integral is calculated in the sense of the Cauchy principal value when $u \in ]-1, 0[$.

Equations (\ref{eq7})-(\ref{eq8}) are retrieved by letting $b \to +\infty$ into Eqs. (\ref{eq16}) and (\ref{eq18}), since $X(a, +\infty, u) = H(a, u)$, $Y(a, +\infty, u) = 0$, $\ksix(a, +\infty, u) = 1/H(a, u)$ and $\ksiy(a, +\infty, u) = 0$ for $u > 0$. Equations (\ref{eq17}) and (\ref{eq19}) reduce to $0=0$ when $b \to +\infty$.

We introduce the moments of order 0 of $X$ and $Y$
\begin{equation}
\label{eq22}
\alpha_{0} (a, b) = \!\! \int_{0}^{1} X(a, b, u) \d u ,\quad \beta_{0} (a, b) = \!\! \int_{0}^{1} Y(a, b, u) \d u,
\end{equation}
and let $u \to +\infty$ into Eqs. (\ref{eq16})-(\ref{eq21}). Since $\ksix(a, b, +\infty) = (a/2)\alpha_{0}(a, b)$ and  $\ksiy(a, b, +\infty) = (a/2)\beta_{0}(a, b)$, one obtains the following relation between the coefficients $\alpha_{0}$ and $\beta_{0}$:
\begin{equation}
\label{eq23}
\left[ 1- \frac{a}{2} \alpha_{0} (a,b) \right]^{2} - \left[ \frac{a}{2} \beta_{0} (a,b) \right]^{2} =1-a,
\end{equation}
and we have further, if $b < +\infty$
\begin{align}
X(a, b, +\infty)  =  Y(a, b, +\infty)
&= \frac{1}{1-a}  \left[ 1 - \frac{a}{2} (\alpha_{0} + \beta_{0})(a,  b) \right],\label{eq24} \\
&=  \frac{1}{\displaystyle 1-\frac{a}{2} (\alpha_{0} - \beta_{0})(a, b)}. \label{eq25}
\end{align}

The generalization of Eqs. (\ref{eq11}) and (\ref{eq15}) in a finite atmosphere are Eqs. (34)-(35) and (58)-(59) of Rutily et al.~\cite{rutily2003a}. We repeat here the two last-mentioned equations, since they will prove useful to the remainder of the present article
\begin{multline}
\label{eq26}
\frac{1}{1-a} \left[ 1 - \frac{a}{2} \alpha_{0} (a, b) \right] = \\  
 1+ R(a)  \left\{ 1 - \ksix [a, b, 1/k(a)] 
 + \ksiy [a, b, 1/k(a)] \exp [- k(a) b] \right\} \\
 + \frac{a}{2} \int_{0}^{1} g(a, v)  [1 - \ksix(a, b, v)
+ \ksiy (a, b, v) \exp (-b/v)] \d v,
\end{multline}
\begin{multline}
\label{eq27}
\frac{1}{1-a} \frac{a}{2} \beta_{0} (a, b) = \\
R(a) \left\{ [1 - \ksix (a, b, 1/k(a))] \exp (- k(a) b) 
+ \ksiy (a, b, 1/k(a)) \right\} \\
+ \frac{a}{2} \int_{0}^{1} g(a, v)  \{ [1-\ksix (a, b, v)] \exp (- b / v) + \ksiy (a, b, v) \} \d v.
\end{multline}

When $b \to +\infty$, Eq. (\ref{eq26}) yields Eq. (\ref{eq15}) again and Eq. (\ref{eq27}) reduces to $0=0$. We have indeed $\beta_0(a, +\infty)=0$, and $\alpha_{0}(a, +\infty) = 2/[1+\sqrt{1-a}]$ due to Eq. (\ref{eq23}).

\section{First method for solving the problem (\ref{eq5})}\label{sec3}

Sobolev~\cite{sobolev1957,sobolev1958} has expressed the resolvent kernel of Eq. (\ref{eq1}) in terms of the associated resolvent function $\phi$, which is solution to the same equation with the free-term $(a/2) E_{1}(\tau)$, viz. 
\begin{equation}
\label{eq28}
\phi (a, b, \tau) = \frac{a}{2} E_{1} (\tau) + \frac{a}{2} \int_{0}^{b} E_{1} (|\tau - \tau\p|) \phi (a, b, \tau\p) \d \tau\p.
\end{equation}

He inferred that the $Q$-function can be expressed in terms of
\begin{equation}
\label{eq29}
\psi (a, b, \tau) = 1 + \int_{0}^{\tau} \phi (a, b, \tau\p) \d \tau\p
\end{equation}
by means of the following formulae: in a semi-infinite medium~\cite{sobolev1957}
\begin{equation}
\label{eq30}
Q (a, \tau) = \frac{1}{\sqrt{1-a}} \;\psi (a, \tau),
\end{equation}
and in a finite one~\cite{sobolev1958}
\begin{equation}
\label{eq31}
Q (a,b, \tau) = \psi (a, b, b) [ \psi (a, b, \tau) + \psi (a, b, b - \tau) - \psi (a, b, b) ],
\end{equation}
where
\begin{equation}
\label{eq32}
\psi (a, b, b) = X (a, b, + \infty) = Y (a, b, + \infty)
\end{equation}
is given by Eqs. (\ref{eq24})-(\ref{eq25}).

The resolvent function has been calculated analytically by Minin~\cite{minin1958} in a semi-infinite medium
\begin{equation}
\label{eq33}
\phi (a, \tau) = 
 \frac{k(a) R(a)}{H [a, 1/k(a)]} \exp [- k(a) \tau] 
+ \frac{a}{2} \int_{0}^{1} \frac{g(a,v)}{H(a, v)} \exp (- \tau / v) \frac{\d v}{v},
\end{equation}
and by Rogovtsov and Samson~\cite{rogovtsov1976} in a finite slab
\begin{multline}
\label{eq34}
\phi (a, b, \tau) = 
k(a) R(a)  \{ [ 1 - \ksix (a, b, 1/k(a))] \exp [- k(a) \tau ] \\ 
\hspace{10em} - \ksiy (a, b, 1/k(a)) \exp [- k(a) (b - \tau) ]  \}  \\
\hspace{-4em} + \frac{a}{2} \int_{0}^{1} g(a, v) \{ [1 - \ksix (a, b, v) ] \exp (- \tau / v) \\
- \ksiy (a, b, v) \exp [- (b - \tau) / v ] \} \frac{\d v}{v}.
\end{multline}

The latter expression does yield the former again when $b \to +\infty$. The functions appearing in the right-hand side have been introduced previously: see Eqs. (\ref{eq12})-(\ref{eq14}) and (\ref{eq20})-(\ref{eq21}).

Substituting these expressions into Eq. (\ref{eq29}), one can perform the integration with respect to $\tau\p$ analytically and simplify the resulting expression of the $\psi$-function with the help of the integral Eqs. (\ref{eq15}) and (\ref{eq26}).
The result is
\begin{equation}
\label{eq35}
\psi (a, \tau) =
\frac{1}{\sqrt{1-a}} - \frac{R(a)}{H (a, 1/k(a))} \exp [- k(a) \tau ] 
 - \frac{a}{2} \int_{0}^{1} \frac{g(a, v)}{H(a, v)} \exp (- \tau / v) \d v
\end{equation}
in a semi-infinite medium, and
\begin{multline}
\label{eq36}
\psi (a, b, \tau) = 
\frac{1}{1-a} \left[ 1 - \frac{a}{2} \alpha_{0} (a, b) \right] \\
 -  R(a) \{ [1 - \ksix (a, b, 1/k(a)) ] \exp [- k(a) \tau ] 
 + \ksiy (a, b, 1/k(a)) \exp [- k(a) (b - \tau) ]  \} \\
 - \frac{a}{2} \int_{0}^{1} g(a, v)  \left\{ [ 1 - \ksix (a, b, v) ] \exp (- \tau / v) 
 + \ksiy (a, b, v) \exp [- (b - \tau ) / v ]  \right\} \d v 
\end{multline}
in the finite case.

We do retrieve $\psi(a, 0) = \psi(a, b, 0) = 1$ in accordance with (\ref{eq29}), due to relations (\ref{eq15}) and (\ref{eq26}). At $\tau = b$, Eq. (\ref{eq36}) reproduces Eq. (\ref{eq32}) owing to (\ref{eq27}).

Expression (\ref{eq35}) of the function $\psi$ was first given by Minin~\cite{minin1958}, apart for an obvious simplification. It appears also in Heaslet et Warming~\cite{heaslet1968}. Its counterpart (\ref{eq36}) in a finite space seems to be new.

The end of the calculation of the $Q$-function is straightforward: in a half-space, one obtains from Eqs. (\ref{eq30}) and (\ref{eq35})
\begin{align}
Q(a,\tau) &= \frac{1}{1-a} [1-\sqrt{1-a}F(a,\tau)], \label{eq37} \\
&= \frac{1}{\sqrt{1-a}} [1+F(a, 0)-F(a,\tau)], \label{eq38}
\end{align}
where
\begin{equation}
\label{eq39}
F(a, \tau) = 
 \frac{R(a)}{H(a, 1/k(a))} \exp [- k(a) \tau ] 
 + \frac{a}{2} \int_{0}^{1} \frac{g(a, v)}{H(a, v)} \exp (- \tau / v) \d v.
\end{equation}

The surface value of $F$ follows from Eq. (\ref{eq15})
\begin{equation}
\label{eq40}
F(a, 0) = \frac{1}{\sqrt{1-a}} - 1,
\end{equation}
which justifies Eq. (\ref{eq38}). Further, we have
\begin{equation}
\label{eq41}
Q(a, 0) = \frac{1}{\sqrt{1-a}}.
\end{equation}
This important result was derived for the first time by Sobolev~\cite{sobolev1957}. It was completed by Minin~\cite{minin1958}, who deduced the first internal expression (\ref{eq37}) of the $Q$-function from his calculation of the resolvent function in a semi-infinite atmosphere.

In a finite medium, we introduce the following functions $F_{+}$ and $F_{-}$:
\begin{multline}
\label{eq42}
F_{\pm} (a, b, \tau) = \\
 R(a)  \{ 1 - \ksix [a, b, 1/k(a)] \pm \ksiy [a, b, 1/k(a)]\} 
 \{ \exp [- k(a) \tau ] \pm \exp [- k(a) (b - \tau )]\}   \\
 + \frac{a}{2} \! \int_{0}^{1} \! g(a, v)  [ 1 - \ksix (a, b, v) \pm \ksiy (a, b, v)  ]
\{ \exp (- \tau / v) \pm \exp [- (b - \tau) / v ]  \} \d v, \!\!\!\!\!
\end{multline}
which yield $F$ again as $b \to +\infty$. Adding and subtracting Eqs. (\ref{eq26}) and (\ref{eq27}), one obtains
\begin{equation}
\label{eq43}
1 - \frac{a}{2} (\alpha_{0} \pm \beta_{0}) (a, b) = (1-a) [1 + F_{\mp} (a, b, 0)],
\end{equation}
so that, from Eq. (\ref{eq23})
\begin{equation}
\label{eq44}
(1-a) [1 + F_{+} (a, b, 0)] [1 + F_{-} (a, b, 0)] = 1.
\end{equation}

The surface values of the functions $F_{\pm}$ can be deduced from Eqs. (\ref{eq43}) and (\ref{eq24})-(\ref{eq25}):
\begin{align}
F_{+}(a, b, 0) &= F_{+}(a, b, b) = \frac{1}{(1-a) X(a, b, +\infty)}-1, \label{eq45} \\
- F_{-}(a, b, 0) &= F_{-}(a, b, b) = 1-X(a, b, +\infty). \label{eq46}
\end{align}

The counterpart of Eqs. (\ref{eq37})-(\ref{eq38}) in a finite slab is, from Eqs. (\ref{eq31}), (\ref{eq32}) and (\ref{eq36})
\begin{align}
Q(a,b,\tau) &= \frac{1}{1-a} [1- (1-a) X(a,b,+\infty)F_{+}(a,b,\tau)], \label{eq47} \\
&= X(a,b,+\infty)[1+F_{+}(a,b, 0)-F_+(a,b,\tau)], \label{eq48}
\end{align}
the second equation resulting from Eq. (\ref{eq45}). We have furthermore
\begin{equation}
\label{eq49}
Q (a, b, 0) = Q (a, b, b) = X (a, b, +\infty).
\end{equation}

From Eqs. (\ref{eq24})-(\ref{eq25}) and (\ref{eq43}), $X(a, b, +\infty)$ is given by 
\begin{equation}
\label{eq50}
X (a, b, +\infty) = 1 + F_{-} (a, b, 0) = \frac{1}{(1-a) [ 1 + F_{+} (a, b, 0)]}.
\end{equation}

Since $X(a, +\infty, +\infty) = 1/\sqrt{1-a}$, Eqs. (\ref{eq47})-(\ref{eq50}) yield Eqs. (\ref{eq37})-(\ref{eq41})  again when $b \to +\infty$.

The surface expression (\ref{eq49}) -- with $X(a, b, +\infty)$ as given by Eqs. (\ref{eq24})-(\ref{eq25}) -- was derived for the first time by Sobolev~\cite{sobolev1958}. The first internal expression (\ref{eq47}) is new, and the second (\ref{eq48}) was given by Danielian~\cite{danielian1983}, using the method summarized in the next section.
\section{Sectionnd method for solving the problem (\ref{eq5})}\label{sec4}
In view of solving the slab albedo problem, Ambartsumian~\cite{ambartsumian1942} introduced the auxiliary function $B = B (a, b, \tau, u)$ solution to the following integral equation:
\begin{equation}
\label{eq51}
B (a, b, \tau, u) = \exp (-\tau / u) 
 + \frac{a}{2} \int_{0}^{b} E_{1} (|\tau - \tau\p|) B(a, b, \tau\p, u) \d \tau\p.
\end{equation}

$(1/4\pi) B(a, b, \tau, u)$ is the mean intensity of the total (direct + diffuse) field as produced at depth $\tau$ by parallel rays incident on the top surface $\tau = 0$ at an angle $\arccos u$ to the inward normal with unit flux per unit area normal to the rays. It is well known~\cite{busbridge1960} that Eq. (\ref{eq51}) has a sense and a unique solution for any $u > 0$, thus for $u \to +\infty$. It coincides then with Eq. (\ref{eq5}), so that one can write
\begin{equation}
\label{eq52}
Q (a, b, \tau) = B (a, b, \tau, +\infty).
\end{equation}

A recent review on the analytical calculation of the $B$-function is in~\cite{rutily2003a}, whose Eqs. (64) and (66) lead to 
\begin{equation}
\label{eq53}
B (a, \tau, +\infty) = H (a, +\infty) \lim_{u \to +\infty} [ u \, \eta (a, \tau, u) ]
\end{equation}
in a semi-infinite medium, and to
\begin{multline}
\label{eq54}
B (a, b, \tau, +\infty) =
X (a, b, +\infty) \lim_{u \to +\infty } [ u \, \etax (a, b, \tau, u) ] \\
- Y (a, b, +\infty) \lim_{u \to +\infty} [ u \, \etay (a, b, \tau, u) ]
\end{multline}
in a finite slab. The limits of the functions $u \eta$, $u \etax$ and $u \etay$ follow immediately from the expressions (69), (74) and (73) of the functions $\eta$, $\etax$ and $\etay$ in~\cite{rutily2003a}.
Substituting these limits into Eqs. (\ref{eq53})-(\ref{eq54}) and using the relations (\ref{eq10}) and (\ref{eq24})-(\ref{eq25}), we retrieve Eqs. (\ref{eq37}) or (\ref{eq47}) depending on whether the atmosphere is semi-infinite or finite.

Danielian~\cite{danielian1983} proceeded that way to calculate the $Q$-function (he denotes by $\bar{N}$): his relation (22), whose left-hand side contains a sign error, coincides with our expression (\ref{eq48}) of the $Q$-function.
\section{Further developments}\label{sec5}
Computing the $Q$-function with the help of Eqs. (\ref{eq37})-(\ref{eq38}) in a half-space, or Eqs. (\ref{eq47}), (\ref{eq48}), (\ref{eq50}) in a finite slab, is not easy for at least two reasons: 1) these expressions contains the indeterminate forms $0\times\infty$ or $\infty - \infty$ as $a\to 1$, since the functions $F$ and $F_+$ diverge when $a\to 1$, while $F_-$ and $X$ remain bounded, 2) in the finite case, the calculation of the functions $F_\pm$ from their definition (\ref{eq42}) is time consuming. It requires the prior calculation of the functions $\ksix$ and $\ksiy$ as defined by Eqs. (\ref{eq20})-(\ref{eq21}), which presupposes the knowledge of the $X$- and $Y$-functions over $[0,1]$.

To get round the first difficulty, we adopt the expressions (\ref{eq38}) and (\ref{eq48}) of the $Q$-function in preference to the expressions (\ref{eq37}) and (\ref{eq47}), and substitute the definitions (\ref{eq39}) and (\ref{eq42}) of the functions $F$ and $F_\pm$ respectively. We obtain
\begin{equation}
\label{eq55a}
Q(a, \tau) = \frac{1}{\sqrt{1-a}} \left[ 1 + \tau f(a, \tau) \right]
\end{equation}
in a half-space, where
\begin{equation}
\label{eq56a}
f(a, \tau) = 
\frac{k(a)R(a)}{H[a,1/k(a)]} \gamma^* \left[ 1,k(a)\tau \right] 
+ \frac{a}{2} \int_0^1 \frac{g(a, v)}{H(a, v)} \, \gamma^*(1,\tau/v) \frac{\d v}{v},
\end{equation}
and
\begin{equation}
\label{eq57a}
Q(a, b, \tau) = X(a, b, +\infty) \left[ 1 + \tau (b - \tau) f(a, b, \tau) \right]
\end{equation}
in a finite slab, where $X(a, b, +\infty)$ is given by the first Eq. (\ref{eq50}), viz.
\begin{multline}
\label{eq58a}
X(a, b, +\infty) = 1 \\ 
+ b \left\{ k(a) R(a) \left[ 1 - \ksix (a, b, 1/k(a)) 
 - \ksiy (a, b, 1/k(a)) \right]  \gamma^* \left[ 1,k(a)\tau \right] \phantom{\frac{\d v}{v}} \right.\\
+ \frac{a}{2} \int_{0}^{1} g(a, v)  \left[ 1 - \ksix (a, b, v) - \ksiy (a, b, v)  \right] 
\left. \gamma^*(1,b/v) \frac{\d v}{v} \right\},
\end{multline}
and
\begin{multline}
\label{eq59a}
f(a, b, \tau) = 
k^2(a) R(a) \left\{ 1 - \ksix [ a, b, 1/k(a) ]  + \ksiy [ a, b, 1/k(a) ] \right\} \\
\hfill \times \gamma^* \left[ 1,k(a)\tau \right] \gamma^* \left[ 1,k(a)(b-\tau) \right] \\
+ \frac{a}{2} \int_{0}^{1} g(a, v)  \left[ 1 - \ksix (a, b, v) + \ksiy (a, b, v)  \right] 
\gamma^*(1,\tau/v) \, \gamma^* [1,(b-\tau)/v] \frac{\d v}{v^2}. 
\end{multline}

We have introduced the function
\begin{equation}
\label{eq60a}
\gamma^*(1,x) = \frac{1}{x} \left( 1 - e^{-x} \right),
\end{equation}
which belongs to a family of classical special functions derived from the incomplete gamma function: see e.g. Abramowitz \& Stegun~\cite{abramowitz1970}, Section 6.5. We note that the values of the function $\gamma^*$ are in the range $[0,1]$ for $x\ge 0$, and that $\gamma^*(1,0)=1$ and $\gamma^*(1,x)\sim 1/x \to 0$ as $x \to +\infty$. A useful expansion of $\gamma^*(1,x)$ for small values of $x$ is
\begin{equation}
\label{eq61a}
\gamma^*(1,x) = e^{-x} \sum_{n=0}^\infty \frac{x^n}{(n \! + \! 1)!}.
\end{equation}

The expression (\ref{eq55a})-(\ref{eq56a}) of the $Q$-function in a semi-infinite atmosphere is our final one, since the $H$-function is easily computable over $[0,1]$ (Section 6). On the other hand, the formulae (\ref{eq57a})-(\ref{eq59a}) involve the functions $\ksix$ and $\ksiy$, which are difficult to compute over $[0,1]$. The problem of calculating these functions is the subject of a wide literature since the fifties~\cite{busbridge1955,mullikin1964,carlstedt1966,rutily1992,danielian1993,danielian1994}. Recently, Rutily et al.~\cite{rutily2003b} have introduced two auxiliary functions $\zetap$ and $\zetam$, depending on parameters $a$, $b$ and variable $u \in \mathbb{R}$, which allow to calculate the functions $X$, $Y$, $\ksix$ and $\ksiy$ in one step. Their calculation algorithm is summarized in the next section. Formulas expressing the functions $\ksix$ and $\ksiy$ are, for $u \geq 0$
\begin{align}
1 - \ksix (a, b, u) = \frac{1}{H(a, u)} \frac{1}{2} (\zetap + \zetam) (a, b, u), \label{eq55} \\
\ksiy (a, b, u) = \frac{1}{H(a, u)} \frac{1}{2} (\zetap - \zetam) (a, b, u). \label{eq56}
\end{align}
Consequently
\begin{equation}
\label{eq57}
1 - \ksix (a, b, u) \pm \ksiy (a, b, u) = \frac{1}{H(a, u)} \zetapm (a, b, u),
\end{equation}
and the expressions (\ref{eq58a}), (\ref{eq59a}) of $X(a, b, +\infty)$ and $f(a, b, \tau)$ read now
\begin{multline}
\label{eq65a}
X(a, b, +\infty) = 
1 + b \left\{ \frac{k(a) R(a)}{H[a,1/k(a)]} 
 \zeta_-[a,b,1/k(a)] \, \gamma^* [1,k(a)b] \right.\\
\left. + \frac{a}{2} \int_{0}^{1} \frac{g(a, v)}{H(a, v)}   \zeta_- (a, b, v) \, \gamma^*(1,b/v) \frac{\d v}{v} \right\},
\end{multline}
\begin{multline}
\label{eq66a}
f(a, b, \tau) = 
\frac{k(a) R(a)}{H[a,1/k(a)]} k(a) \zeta_+ [a, b, 1/k(a)] \\ 
\times  \gamma^*[1,k(a)\tau] \, \gamma^*[1,k(a)(b-\tau)] \\
+ \frac{a}{2} \int_{0}^{1} \frac{g(a, v)}{H(a, v)} \zeta_+ (a, b, v) \, \gamma^*(1,\tau/v) \, \gamma^*[1,(b-\tau)/v] \frac{\d v}{v^2}.
\end{multline}

When $b \to +\infty$, these formulae yield $X(a,+\infty,+\infty) = 1/\sqrt{1-a}$ and \\$f(a,+\infty,\tau) = f(a,\tau)$ [given by (\ref{eq56a})] as it should, since $\zeta_\pm(a,b,u) \to 1$ from Eqs. (\ref{eq62})-(\ref{eq66}) below, and $b \gamma^*(1,\alpha b) \to 1/\alpha$ from Eq.~(\ref{eq60a}) ($\alpha > 0$).
\section{Numerical calculations}\label{sec6}
The numerical evaluation of the $Q$-function with the help of relations (\ref{eq55a})-(\ref{eq56a}) and (\ref{eq57a}), (\ref{eq65a}), (\ref{eq66a}) requires 
the previous calculation of four auxiliary functions: $k(a)$, $H(a, u)$, $\zeta_{+}(a, b, u)$ and $\zeta_{-}(a, b, u)$ for $u \in [0,1]$ and at $u = 1/k(a)$.
The functions $R(a)$, $g(a, v)$ and $\gamma^*(1,x)$ are defined by Eqs. (\ref{eq13}), (\ref{eq14}) and (\ref{eq60a}).

\subsection{Coefficient $k(a)$}
We used the analytical expression of $k(a)$ first given by Carlstedt and Mullikin~\cite{carlstedt1966}, viz.
\begin{equation}
\label{eq59}
k (a) = \sqrt{1-a} \, \exp \left[ \int_{0}^{1} \theta (a, v) \d v / v \right],
\end{equation}
where
\begin{equation}
\label{eq60}
\theta (a, v) = \frac{1}{\pi} \arctan \left[ \frac{\pi}{2} \frac{a v}{T(a,v)} \right] \qquad (0 \leq v < 1).
\end{equation}

Continuous values on $[0,\pi]$ of the arctan function are used in the above definition of the $\theta$-function, i.e., the branch is not the principal one. The resulting function $\arctan(x/y)$ is usually denoted by $\mathrm{ATAN2}(x,y)$. With this choice, the function $v \to \theta(a, v)$ is continuous from $[0,1[$ to $[0,1[$, although $T(a, v)$ vanishes once in the interval $[0,1[$.
\subsection{Function $H(a, u)$}
There are many analytical expressions of this function in the literature, including the following one~\cite{mullikin1964}
\begin{equation}
\label{eq61}
H (a, u) = \frac{1 + u}{1 + u \, k(a)} \exp \left[ u \int_{0}^{1} \theta (a, v) \frac{\d v}{v(v + u)} \right] \qquad (u \geq 0)
\end{equation}
we have selected.
\subsection{Functions $\zetapm(a, b, u)$}
The definition and the method of calculation of these functions are explained in two papers in preparation~\cite{rutily2003,chevallier2003a}.
Here, we just give the useful formulas, with no justification.

For any $u \geq 0$, one has
\begin{equation}
\label{eq62}
\zetapm (a, b, u) = \rho_{\pm} (a, b, u) \pm M_{\pm} (a, b) \sigma_{\pm} (a, b, u),
\end{equation}
where $\rho_{\pm}$ and $\sigma_{\pm}$ are solution to the integral equations
\begin{equation}
\label{eq63}
\rho_{\pm} (a, b, u) =  1 
 \pm \frac{a}{2} u \int_{0}^{1} \frac{g(a, v)}{H^{2} (a, v)}
\exp (-b / v) \rho_{\pm} (a, b, v) \frac{\d v}{v + u} \, , 
\end{equation}
\begin{equation}
\label{eq64}
\sigma_{\pm} (a, b, u) =
\frac{2 k(a) u}{1+ k(a) u} 
 \pm \frac{a}{2} u \int_{0}^{1} 
\frac{g(a,v)}{H^{2} (a, v)} \exp (-b / v) \sigma_{\pm} (a, b, v) \frac{\d v}{v + u} \, ,
\end{equation}
which are of Fredholm type on $[0,1]$. Once they have been solved on this interval, these equations define the functions $u \to \rho_{\pm}(a, b, u)$ and $u \to \sigma_{\pm}(a, b, u)$ on $[0, +\infty[$, in particular at $u = 1/k(a)$. This yields the coefficient $M_{\pm}(a, b)$ appearing in Eq. (\ref{eq62}), since
\begin{equation}
\label{eq65}
M_{\pm} (a, b) = \frac{q (a, b) \rho_{\pm} [a, b, 1/k(a)]}
{1 \mp q (a, b) \sigma_{\pm} [a, b, 1/k(a)] },
\end{equation}
where
\begin{equation}
\label{eq66}
q (a, b) = \frac{1}{2} R(a) \frac{\exp [ - k(a) b ]}{H^{2} [a, 1/k(a)]}.
\end{equation}

Calculating the functions $\zeta_{\pm}$ thus requires the numerical solution of Eqs. (\ref{eq63}) and (\ref{eq64}).
Actually, it is proved in Ref.~\cite{rutily2003} that the solution of Eqs. (\ref{eq64}) can be reduced to that of Eqs. (\ref{eq63}), i.e., the functions $\sigma_{\pm}$ are analytically expressible in terms of the  functions $\rho_{\pm}$.
It follows that Eqs. (\ref{eq63}) are the only ones to be solved numerically.
Details are not given here, since there is no inconvenience, at the level of accuracy, to solve both sets of equations numerically.
Indeed, the free term and kernel of Eqs. (\ref{eq63})-(\ref{eq64}) are regular over $[0, 1]$ and $[0,1]\times [0,1]$ respectively.
Their solution $\rho_{\pm}$ and $\sigma_{\pm}$ are regular and smooth over $[0,1]$, i.e., not only continuous, but derivable everywhere in $[0,1]$, including at 0.

We note that our algorithm for calculating the functions $1 - \ksix \pm \ksiy$ with the help of Eqs. (\ref{eq57}) and (\ref{eq62})-(\ref{eq66}) was already introduced, apart from a few details, by Rutily~\cite{rutily1992} and Danielian~\cite{danielian1993,danielian1994}.

Every integral in the above expressions has been calculated using the routines of the NAG (Numerical Algorithms Group) Fortran Library (D01AHF and D01AJF). We also took the routine for solving the Fredholm integral equations (\ref{eq63})-(\ref{eq64}) from this library (D05ABF).
We have collected in~\cite{chevallier2003a} some arguments showing that these integral equations are currently solved with an accuracy better than $10^{-11}$ for any value of $a$ and $b$.

For given functions $H(a, v)$ and $\zeta_{\pm}(a, b, v)$, the numerical evaluation of the integrals in Eqs. (\ref{eq56a}), (\ref{eq65a}) and (\ref{eq66a}) is difficult in two circumstances: ({\em i\/}) in a weakly scattering medium ($a$ close to 0), due to the rapid variation of the function $v \to g(a, v)$ in a neighbourhood of $v = 1$, ({\em ii\/}) when $\tau$ tends to 0 on the right or to $b$ on the left, because of the sharp maximum of the integrand close to $v = \tau$ or $v = b - \tau$ (resulting from the presence of the exponentials).
These difficulties can be overcome thanks to two changes of variable.
The first one ({\em i\/}) by choosing $t = [1 - \ln(1-v)]^{-1}$ as a new variable, the regularizing effect of this transformation being explained in details in the Section 8.1.3 of~\cite{rutily1992}.
The second difficulty ({\em ii\/}) is tractable by making the change of variable $x = [1 - (\ln v)^{-1}]^{-1}$, since it substantially reduces the strong variations of the function $v \to \exp(-\tau/v)$ in a neighbourhood of $v = 0$ when $\tau \to 0^{+}$ \cite{bergeat2000}.
When $\tau \to b^{-}$, we come down to the preceding case by remarking that $f(a, b, \tau) = f(a, b, b - \tau)$.

\begin{figure}[htb]
\begin{center}
\resizebox{0.7\hsize}{!}{\includegraphics[angle=-90]{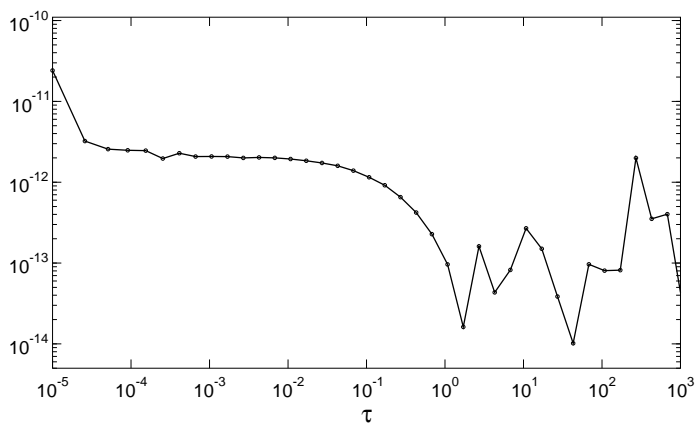}}
\end{center}
\caption{Residuals of Eq. (\ref{eq5}) as a function of $\tau$, showing the accuracy of our solution for $1-a = 10^{-4}$ and $b = 2000$. These residuals are symmetric about the $\tau$-mid-plane.}
\label{fig1}
\end{figure}

With these precautions, the integrals appearing in Eqs. (\ref{eq56a}), (\ref{eq65a}) and (\ref{eq66a}) can be calculated with an accuracy better than $10^{-11}$ using the routines of the NAG library.
Finally, we may safely conjecture that the problem (\ref{eq5}) has been solved with an accuracy better than $10^{-10}$, this estimation being corroborated by the curve of residuals of Fig.~\ref{fig1}.
Residuals are the relative difference between the left- and right-hand sides of Eq. (\ref{eq5}), calculated for any value of $\tau$ in $[0, b]$.
\subsection{The case $a \to 1$}
Until now, the conservative case $a = 1$ has been excluded, since most of the preceding results lead to indeterminate forms when $a=1$.
This situation generates important roundoff errors when $a\to 1$, which are incompatible with the required accuracy.
To remove this difficulty, it is appropriate to introduce asymptotic expansions of the functions $k(a)$, $H(a, u)$, ... for $a\to 1$.

The following expansion of $k(a)$ is given by Case et al.~\cite{case1953}:
\begin{multline}
\label{eq67}
k (a) =
\sqrt{3(1-a)}  \left[ 1 - \frac{2}{5} (1-a) - \frac{2^{2} \times 3}{5^{2} \times 7} (1-a)^{2} - \frac{ 2}{5^{3}} (1-a)^{3} \right. \\
\left. \phantom{\frac{2^2}{5^2}} + \frac{2\times 83}{5^{3} \times 7^{2} \times 11} (1-a)^{4} + ...  \right] \quad (a \to 1).
\end{multline}

The calculation of the $H$-function from its expression (\ref{eq61}) does not raise any difficulty when $a\to 1$, except at $u = 1/k(a)$. Actually, Eq. (\ref{eq15}) clarifies the behaviour of the ratio $k(a) R(a) / H[a,1/k(a)]$ appearing in Eqs. (\ref{eq56a}), (\ref{eq65a}) and (\ref{eq66a}), as $a\to 1$. We have in particular
\begin{equation}
\label{eq68}
\frac{k(a) R(a)}{H[a,1/k(a)]}\sim \sqrt{3} \quad {\rm as} \quad a\to 1.
\end{equation}

As a result, the $f$-function defined by Eq. (\ref{eq56a}) tends to $f(1,\tau)=\sqrt{3} + \left[ \sqrt{3} q(\tau) -1 \right] / \tau$ as $a\to 1$, where
\begin{equation}
\label{eq70}
q(\tau)=\frac{1}{\sqrt{3}} \left[ 1+\frac{\tau}{2}\int_0^1 \frac{g(1,v)}{H(1,v)} \, \gamma^* ( 1,\tau/v) \frac{\d v}{v} \right]
\end{equation}
is Hopf's function~\cite[p. 139]{ivanov1973}. This function increases from $q(0)=1/\sqrt{3}$ to $q(\infty)=0.71044608959876$; it is tabulated e.g. in~\cite[p. 141]{ivanov1973}.

We conclude that in a semi-infinite atmosphere
\begin{equation}
\label{eq78a}
Q(a,\tau) \sim \sqrt{3/(1-a)} \left[ \tau + q(\tau) \right] \quad (a\to 1).
\end{equation}

In the finite case, there is no difficulty to compute the functions $\rho_{\pm}$ and $\sigma_{\pm}$ from Eqs. (\ref{eq63})-(\ref{eq64}) as $a\to 1$. On the other hand, the coefficient $M_{+}(a, b)$ diverges like $1/k(a)$ as $a\to 1$, while $M_{-}(a, b)$ remains bounded. Since $\sigma_{+}(a, b, u) \propto k(a)$ for $u \neq 1/k(a)$, $\zetap(a, b, u)$ as defined by (\ref{eq62}) is bounded when $u \neq 1/k(a)$, and diverges like $1/k(a)$ for $u = 1/k(a)$. Indeed, putting $u=1/k(a)$ and $a\to 1$ into Eq. (\ref{eq57}) yields
\begin{equation}
\label{eq72}
k(a) \zeta_+[a, b,1/k(a)]\sim \frac{\sqrt{3}}{2} \beta_0(1,b) \quad {\rm as} \quad a\to 1,
\end{equation}
where $\beta_0(1,b)$ is the value at $a=1$ of the $\beta_0$-coefficient defined by the second Eq. (\ref{eq22}). On the other hand, $\zetam(a, b, u)$ remains bounded as $a\to 1$ whatever $u\geq 0$.

It follows from Eqs. (\ref{eq23})-(\ref{eq25}) that
\begin{equation}
\label{eq80a}
X(1,b,+\infty) = \frac{1}{\beta_0(1,b)} .
\end{equation}
Moreover, taking Eqs. (\ref{eq68}) and (\ref{eq72}) into account, Eq. (\ref{eq66a}) reads at $a=1$
\begin{multline}
\label{eq81a}
f(1,b,\tau) = \frac{3}{2} \, \beta_0(1,b) \\ 
+ \frac{1}{2} \int_0^1 \frac{g(1,v)}{H(1,v)} \zeta_+(1,b,v) \, \gamma^* (1, \tau/v) \, \gamma^* [1, (b - \tau)/v] \frac{\d v}{v^2} .
\end{multline}

We conclude that the $Q$-function is bounded as $a\to 1$, with expression given by Eqs. (\ref{eq57a}), (\ref{eq80a}) and (\ref{eq81a}) at $a=1$, viz.
\begin{multline}
\label{eq82a}
Q(1,b,\tau) = \frac{3}{2} \tau (b - \tau) 
+ \frac{1}{\beta_0(1,b)} \left\{ 1 + \tau (b - \tau) \phantom{\frac{1}{2}\int_0^1\frac{\d v}{v^2}} \right. \\ 
\left. \times \frac{1}{2}  \int_0^1 \frac{g(1,v)}{H(1,v)} \zeta_+(1,b,v) \, \gamma^* (1, \tau/v) \, \gamma^* [1, (b - \tau)/v] \frac{\d v}{v^2} \right\} .\end{multline}

This is an alternative form of Danielian's expression for $a=1$~\cite[Eq. (23)]{danielian1983}. The coefficient $\beta_0(1,b)$ can be accurately calculated using the following formula, given here with no proof:
\begin{equation}
\label{eq75}
\beta_0(1, b)= \frac{1}{\sqrt{3}}\;\frac{1}{[b/2+q(\infty)]\rho_{-,0}(1,b)+\rho_{-,1}(1,b)},
\end{equation}
where we have introduced the coefficients
\begin{equation}
\label{eq76}
\rho_{\pm,n} (a, b) =  \delta_{n,0} 
\pm \frac{a}{2} \int_{0}^{1} \frac{g(a, v)}{H^{2} (a, v)}
\exp (-b / v) \rho_{\pm} (a, b, v) v^n \d v \quad (n\geq 0) 
\end{equation}
($\delta_{n,0}$ = Kronecker symbol).
\section{Tables of the function $S=(1-a) Q$}\label{sec7}
Numerical data concerning the $Q$-function are few. We are aware of some values published by Sobolev and Minin~\cite{sobolev1961}, and of far more detailed tables given by Buell et al.~\cite{buell1971} using the invariant embedding method.
In view of checking their five-digit tables, we have computed the $Q$-function for those values of $a$ and $b$ they have selected (Table~\ref{tab_buell}). We got a poor agreement for small values of $\tau$, most of time 3 digits in agreement only.
\begin{table}
\caption{Values of $Q(a,b,\tau)$ for $a$, $b$ and $\tau$ taken from Tables 1-4 of Buell et al.~\cite{buell1971}, where they are denoted as $\lambda$, $x$ and $t$ respectively.}
\label{tab_buell}
\centering
\begin{tabular}{ll}
\begin{tabular}[t]{lll}
\hline
\noalign{\smallskip}
$a, b$ & $\tau$ & $Q(a, b, \tau)$ \\
\noalign{\smallskip}
\hline\noalign{\smallskip}
0.2, 1 & 0     & 1.0964212391 \\
       & 0.005 & 1.0994952541 \\
       & 0.01  & 1.1018130569 \\
       & 0.5   & 1.1542831599 \\
\noalign{\smallskip}
\hline\noalign{\smallskip}
0.2, 5 & 0   & 1.1178464365 \\
       & 0.5 & 1.2013059145 \\
       & 1   & 1.2259902468 \\
       & 1.5 & 1.2368227109 \\
       & 2   & 1.2415886078 \\
       & 2.5 & 1.2429482333 \\
\noalign{\smallskip}
\hline\noalign{\smallskip}
0.6, 1 & 0   & 1.3962004895 \\
       & 0.1 & 1.5178886319 \\
       & 0.2 & 1.5816080857 \\
       & 0.3 & 1.6213474973 \\
       & 0.4 & 1.6435798894 \\
       & 0.5 & 1.6507643063 \\
\noalign{\smallskip}
\hline\noalign{\smallskip}
0.6, 5 & 0   & 1.5769290982 \\
       & 1.5 & 2.3199770925 \\
       & 2   & 2.3672862917 \\
       & 2.5 & 2.3816047458 \\
\noalign{\smallskip}
\hline
\end{tabular}
&
\begin{tabular}[t]{lll}
\hline
\noalign{\smallskip}
$a, b$ & $\tau$ & $Q(a, b, \tau)$ \\
\noalign{\smallskip}
\hline\noalign{\smallskip}
0.9, 4 & 0   & 2.8413250008 \\
       & 0.5 & 4.5154025141 \\
       & 1   & 5.3716006806 \\
       & 2   & 5.9854344902 \\
\noalign{\smallskip}
\hline\noalign{\smallskip}
0.9, 10 & 0  & 3.1478960008 \\
        & 0.5 & 5.1418113055 \\
        & 3   & 8.6081352918 \\
        & 5   & 9.1352617263 \\
\noalign{\smallskip}
\hline\noalign{\smallskip}
1, 1 & 0   & 2.0673668738 \\
     & 0.1 & 2.4025482660 \\
     & 0.2 & 2.5919961827 \\
     & 0.3 & 2.7148715436 \\
     & 0.4 & 2.7852328124 \\
     & 0.5 & 2.8082221832 \\
\noalign{\smallskip}
\hline\noalign{\smallskip}
1, 10 & 0 & 9.8907816365 \\
      & 1  & 25.586442568 \\
      & 2  & 36.264849535 \\
      & 3  & 43.801265888 \\
      & 4  & 48.310055291 \\
      & 5  & 49.811793274 \\
\noalign{\smallskip}
\hline
\end{tabular}
\\
\end{tabular}
\end{table}

It is clear that the function $Q$ is appropriate for solving analytically the problem (1) in a homogeneous and isothermal atmosphere, the solution being given by (4).
From a numerical point of view, the one drawback of this function is that it diverges like $1/\sqrt{1-a}$ when $a\to 1$, $b \to +\infty$ and $\tau \to 0$.
That's why we chose to tabulate the function $S=(1-a) Q$ solution to Eq. (\ref{eq6}).
It is the function that is generally calculated as a benchmark solution to the problem (\ref{eq1}). 

The surface values of the $S$-function follow from Eqs. (\ref{eq41}) and (\ref{eq49}), which lead to the famous $\sqrt{1-a}$-law in a half-space
\begin{equation}
\label{eq77}
S(a, 0)=\sqrt{1-a},
\end{equation}
and to
\begin{equation}
\label{eq78}
S(a, b,0)=S(a, b, b)=(1-a) X(a,b,+\infty)
\end{equation}
in a finite slab.
Table~\ref{tab_surface} contains surface values $S(a, b, 0)$ for a wide range of parameters $a$ and $b$, which provides a good test for codes solving the transfer equation in a finite slab.
\begin{table}
\caption{Surface values $S(a,b,0)$ for a wide range of parameters $a$ and $b$.}
\label{tab_surface}
\centering
\begin{tabular}{llll}
\hline
\noalign{\smallskip}
$b$ & $1-a = 10^{-10}$ & $1-a = 10^{-8}$ & $1-a = 10^{-6}$ \\
\noalign{\smallskip}
\hline\noalign{\smallskip}
2              & 2.9549172423E-10 & 2.9549171567E-08 & 2.9549086009E-06  \\
20             & 1.8551036585E-09 & 1.8551015456E-07 & 1.8548902847E-05  \\
200            & 1.7443543255E-08 & 1.7441791865E-06 & 1.7268758494E-04  \\
2000           & 1.7331077826E-07 & 1.7161299590E-05 & 9.3944242810E-04  \\
$2\times 10^4$ & 1.7150550856E-06 & 9.3931230259E-05 & 1.0000000000E-03  \\
$2\times 10^6$ & 1.0000000000E-05 & 1.0000000000E-04 & 1.0000000000E-03  \\
$2\times 10^8$ & 1.0000000000E-05 & 1.0000000000E-04 & 1.0000000000E-03  \\
\noalign{\smallskip}
\hline
\noalign{\smallskip}
$b$ &  $1-a = 10^{-4}$ & $1-a = 0.01$ & $a = 0.9$ \\
\noalign{\smallskip}
\hline\noalign{\smallskip}
2                & 2.9540533187E-04  & 2.8714541813E-02 & 2.3161893202E-01 \\
20               & 1.8340522066E-03  & 9.5164307616E-02 & 3.1622023495E-01 \\
200              & 9.4072188464E-03  & 1.0000000000E-01 & 3.1622776602E-01 \\
2000             & 1.0000000000E-02  & 1.0000000000E-01 & 3.1622776602E-01 \\
$2\times 10^4$   & 1.0000000000E-02  & 1.0000000000E-01 & 3.1622776602E-01 \\
$2\times 10^6$   & 1.0000000000E-02  & 1.0000000000E-01 & 3.1622776602E-01 \\
$2\times 10^8$   & 1.0000000000E-02  & 1.0000000000E-01 & 3.1622776602E-01 \\
\noalign{\smallskip}
\hline\noalign{\smallskip}
$b$  & $a = 0.5$ & $a = 0.1$ & $a = 10^{-3}$ \\
\noalign{\smallskip}
\hline\noalign{\smallskip}
2               & 6.8787981968E-01 & 9.4657359383E-01  & 9.9948108678E-01 \\
20              & 7.0710678074E-01 & 9.4868329804E-01  & 9.9949987494E-01 \\
200             & 7.0710678119E-01 & 9.4868329805E-01  & 9.9949987494E-01 \\
2000            & 7.0710678119E-01 & 9.4868329805E-01  & 9.9949987494E-01 \\
$2\times 10^4$  & 7.0710678119E-01 & 9.4868329805E-01  & 9.9949987494E-01 \\
$2\times 10^6$  & 7.0710678119E-01 & 9.4868329805E-01  & 9.9949987494E-01 \\
$2\times 10^8$  & 7.0710678119E-01 & 9.4868329805E-01  & 9.9949987494E-01 \\
\noalign{\smallskip}
\hline
\end{tabular}
\end{table}

Within the atmosphere, the $S$-function is calculated by multiplying both sides of Eqs. (\ref{eq55a}) and (\ref{eq57a}) by $1-a$.
We obtain
\begin{equation}
S(a, \tau) = \sqrt{1-a} \left[ 1 + \tau f(a, \tau) \right]
\end{equation}
in a semi-infinite atmosphere, and
\begin{equation}
S(a, b, \tau) = (1-a) X(a, b, +\infty) \left[ 1 + \tau (b - \tau) f(a, b, \tau) \right]
\end{equation}
in a finite slab.

The $S$-function has been calculated for the following couples ($a$, $b$): (0.5, 2), (0.99, 20), ($1-10^{-4}$, $2\times 10^{3}$) and ($1-10^{-8}$, $2\times 10^{8}$) (Table~\ref{tab_s}).
The first value represents an ``easy'' case, the second, the third and the fourth ones are typical of a stellar continuum, an ``average'' line and a strong line respectively.
All data are given with eleven figures, which we hope to be significant in view of the residuals of Fig.~\ref{fig1}.
\begin{table}
\caption{$S(a,b, \tau)$ as a function of $\tau$ for several couples ($a$, $b$).}
\label{tab_s}
\centering
\begin{tabular}{ll}
\begin{tabular}[t]{ll}
\hline\noalign{\smallskip}
\multicolumn{2}{l}{$a = 0.5$, $b = 2$}\\
\noalign{\smallskip}
\hline\noalign{\smallskip}
$\tau$ & $S(a,b, \tau)$ \\
\noalign{\smallskip}
\hline\noalign{\smallskip}
0                  & 6.8787981968E-01 \\
$10^{-4}$          & 6.8805116359E-01 \\
$5 \times 10^{-4}$ & 6.8859830701E-01 \\
$10^{-3}$          & 6.8919789714E-01 \\
$5 \times 10^{-3}$ & 6.9309464759E-01 \\
0.01               & 6.9713223336E-01 \\
0.05               & 7.2064553226E-01 \\
0.1                & 7.4199582863E-01 \\
0.5                & 8.2863857712E-01 \\
1                  & 8.5687200196E-01 \\
\noalign{\smallskip}
\hline
\hline\noalign{\smallskip}
\multicolumn{2}{l}{$a = 1-10^{-4}$, $b = 2000$}\\
\noalign{\smallskip}
\hline\noalign{\smallskip}
$\tau$ & $S(a,b, \tau)$ \\
\noalign{\smallskip}
\hline\noalign{\smallskip}
0         & 1.0000000000E-02 \\
$10^{-4}$ & 1.0005867102E-02 \\
$10^{-3}$ & 1.0047210816E-02 \\
0.01      & 1.0359952897E-02 \\
0.1       & 1.2588036202E-02 \\
0.5       & 2.0308299268E-02 \\
1         & 2.9070093404E-02 \\
5         & 9.4251967156E-02 \\
10        & 1.6938939655E-01 \\
100       & 8.2524627065E-01 \\
1000      & 9.9999994061E-01 \\
\noalign{\smallskip}
\hline
\end{tabular}
&
\begin{tabular}[t]{ll}
\hline\noalign{\smallskip}
\multicolumn{2}{l}{$a = 0.99$, $b = 20$}\\
\noalign{\smallskip}
\hline\noalign{\smallskip}
$\tau$ & $S(a,b, \tau)$ \\
\noalign{\smallskip}
\hline\noalign{\smallskip}
0                  & 9.5164307616E-02 \\
$10^{-4}$          & 9.5218140055E-02 \\
$10^{-3}$          & 9.5594616111E-02 \\
0.01               & 9.8408033401E-02 \\
0.1                & 1.1789213897E-01 \\
0.5                & 1.8112719045E-01 \\
1                  & 2.4654055042E-01 \\
2                  & 3.5613573437E-01 \\
5                  & 5.7511174272E-01 \\
10                 & 6.9556824233E-01 \\
\noalign{\smallskip}
\hline
\hline\noalign{\smallskip}
\multicolumn{2}{l}{$a = 1-10^{-8}$, $b = 2\times 10^8$}\\
\noalign{\smallskip}
\hline\noalign{\smallskip}
$\tau$ & $S(a,b, \tau)$ \\
\noalign{\smallskip}
\hline\noalign{\smallskip}
0         & 1.0000000000E-04 \\
$10^{-4}$ & 1.0005884722E-04 \\
$10^{-3}$ & 1.0047386240E-04 \\
0.01      & 1.0361724723E-04 \\
0.1       & 1.2607724794E-04 \\
1         & 2.9416052346E-04 \\
10        & 1.8533929890E-03 \\
100       & 1.7292311586E-02 \\
1000      & 1.5913835240E-01 \\
$10^4$    & 8.2310056337E-01 \\
$10^8$    & 1.0000000000E-00 \\
\noalign{\smallskip}
\hline
\end{tabular}
\\
\end{tabular}
\end{table}
\section{Conclusion}\label{sec8}
Many codes for solving the radiative transfer equation in a stellar atmosphere make use of the idealized problems (\ref{eq5}) or (\ref{eq6}) as a computational test.
The simplest test is provided by the surface value of the solution $S$ to (\ref{eq6}), which is close to $\sqrt{1-a}$ in a slab of great optical thickness.
This well-known result does not turn a rich literature on the subject to best account, which has motivated the writing of the present paper.

From an analytical point of view, we have retrieved the solution to Eq. (\ref{eq5}) in a semi-infinite atmosphere, which can be attributed to Sobolev~\cite{sobolev1957} and Minin~\cite{minin1958}.
In a finite atmosphere, we have completed the solution given by Rogovtsov and Samson~\cite{rogovtsov1976} and Danielian~\cite{danielian1983}.
Our approach is classical up to and including Section \ref{sec4}, as it was developed essentially in the countries of the ex-USSR since the end of the fifties.

In Section \ref{sec5}, the solution to Eq. (\ref{eq5}) has been expressed in terms of two non classical auxiliary functions, denoted by $\zetap$ and $\zetam$, which are supposed to replace the functions $X$, $Y$, $\ksix$ and $\ksiy$.
Actually, there are some evidences that these four functions are not the best intermediate step for solving the radiative transfer equation within a finite slab.
In a paper in preparation~\cite{ahues2003}, we will describe a more direct solution to Eq. (\ref{eq5}), which does not involve the auxiliary functions $\phi$, $\psi$, $B$, etc.

Finally, accurate tables of the surface and internal values of the function $S=(1-a) Q$ are given in this article. We have already taken advantage of these tables in a recent article testing the accuracy of the ALI code, widely used in stellar atmospheres modelling~\cite{chevallier2003b}.
%

\end{document}